\documentclass[aps,prl,twocolumn,amsmath,amssymb,nofootinbib,superscriptaddress,floatfix]{revtex4}
\usepackage{color}
\usepackage{graphicx}
\usepackage{amsmath}
\usepackage{amssymb}
\usepackage{epsfig}
\usepackage{wasysym}
\usepackage{bbm}
\newcommand{\Fig}[1]{Fig.~\ref{#1}}

\newcommand \ti[1]{}

\usepackage{multirow}
\renewcommand{\narrowtext}{\begin{multicols}{2} \global\columnwidth20.5pc}
\renewcommand{\v}[1]{{\bf #1}}

\newcommand{\s}{{\sigma}}

\def\be{\begin{eqnarray}}
\def\ee{\end{eqnarray}}

\newcommand{\Eq}[1]{Eq.~(\ref{#1})}

\newcommand{\ra}{\rightarrow}

\newcommand{\e}{\epsilon}

\begin{document}

\title{Fermiology, Orbital order, Orbital Fluctuation and Cooper Pairing in Iron-based Superconductors}
\author{Fan Yang}
\affiliation{School of Physics, Beijing Institute of Technology, Beijing 100081, P.R.China}
\author{Fa Wang}
\affiliation{
International Center for Quantum Materials and School of Physics, Peking
University, Beijing 100871, China}
\author{Dung-Hai Lee}
\affiliation{
Department of Physics,University of California at Berkeley,
Berkeley, CA 94720, USA}
\affiliation{Materials Sciences Division,
Lawrence Berkeley National Laboratory, Berkeley, CA 94720, USA}

\begin{abstract}
We address two important issues that arise in recent studies of iron-based superconductivity. (1) Why are the T$_c$ of A$_x$Fe$_{2-y}$Se$_2$ and the single unit cell FeSe on SrTiO$_3$ so high despite both only have electron pockets? (2) What (if any) are the effects of orbital order and orbital fluctuation on the Cooper pairing. Our conclusions are summarized in the third paragraph of the paper.
\end{abstract}

\maketitle

 The discovery of A$_x$Fe$_{2-y}$Se$_2$\cite{guo} (T$_c^{max}=48K$, under pressure\cite{sun}) and single unit cell FeSe on SrTiO$_3$ (FeSe/STO)\cite{xue} (T$_c^{max}$=65K, determined by angle-resolved photoemission spectroscopy (ARPES)\cite{xj}), stirred up a new wave of excitement in  iron-based superconductors (FeSCs) research.  In ARPES studies it is found, at ambient pressure, both systems have no hole pocket\cite{donglai, xj}. 
 Because it is often perceived that the scattering between the electron and hole pockets are important for both antiferromagnetism and Cooper pairing\cite{mazin}, this becomes an issue. 

On a different front, recently many experimental evidences point to the fact that FeSCs have a tendency to become electronically ``nematic''\cite{davis,chu1,yi,matzuda,chu2}. For example, through magnetic torque measurement Ref.\cite{matzuda} reported a phase diagram for BaFe$_2$(As$_{1-x}$P$_x$)$_2$ where the superconducting dome is enclosed by a non-magnetic electronic nematic phase. In addition, Ref.\cite{chu2} reported a divergent ``nematic susceptibility'' in Ba(Fe$_{1-x}$Co$_x$)$_2$As$_2$ close to the $x$ value at which T$_c^{max}$ occurs. In addition to these, an ARPES experiment by Yi {\it et al.}\cite{yi} established the tie between electronic nematicity and the $d_{xz}, d_{yz}$ orbital ordering. These experiments naturally raise the question: what role  (if any) do orbital order or orbital fluctuation play in Cooper pairing?

The purpose of this paper is to address the above two questions. Our conclusions are summarized as follows.
(1)  Hole pockets introduce frustration in Cooper pairing (a concept we shall discuss later).  To a large extent this is due to the existence of band vorticity around the hole fermi surface (see later). Removing the hole pockets releases pairing frustration, experimentally it is found this does not weaken the antiferromagnetic (AFM) correlation\cite{jun}. This makes Cooper pairing in A$_x$Fe$_{2-y}$Se$_2$ and FeSe/STO stronger, hence higher T$_c$. Of course in FeSe/STO substrate screening can further enhance T$_c$\cite{fese}. (2) Orbital fluctuation has negligible effect on Cooper pairing while static orbital order can have large effect.  (3) AFM fluctuation can still be the primary cause of Cooper pairing, but AFM is most likely due to local correlation not fermi surface nesting. (4) The interpocket hybridization tends to favor {\it in phase} s-wave pairing in A$_x$Fe$_{2-y}$Se$_2$ and FeSe/STO.
\\

The above conclusions are reached  by using 
an effective Hamiltonian approach. 
In this approach we write down a low energy Hamiltonian to capture the system's tendencies toward (1) stripe antiferromagnetism\cite{afm1,afm2,afm3} (2) superconducting pairing\cite{hosono,zhao}, and (3) $d_{xz}/d_{yz}$ orbital ordering\cite{davis,chu1,chu2,matzuda,yi,XueQK}. 
 The above tendencies are not only established experimentally, but also found theoretically\cite{mazin, Dong,kuroki, fa, ku}. To a great extent, the derivation of the effective Hamiltonian has been achieved by the functional renormalization group (FRG) study in Ref.\cite{zhai}. It's conclusion has been checked by variational Monte-Carlo (VMC) calculation\cite{fan}. The last statement is significant because FeSCs are by no means weakly coupled systems\cite{basov}.

Our effective Hamiltonian is given by:
\begin{eqnarray}
H_{\rm eff}=&&\sum_{\v k,\alpha,\s}\e_{\v k\alpha}n_{\v k\alpha\s}+\sum_{\v k,\alpha,\s}\sum_{\v q,\beta,\tau} F_{\v k\alpha\s}^{\v q\beta\tau}n_{\v k\alpha\s}n_{\v q\beta\tau}+\nonumber\\&&\sum_{ij}J_{ij}\v S_i\cdot\v S_j+V\sum_i n_{i,xz}n_{i,yz}.
\label{eff}
\end{eqnarray}
The first term is the bandstructure ($\alpha,\beta=$ band indices, $\s,\tau=$ spin indices). The second term is a multi-band version of the Fermi liquid interaction. 
 In the third term $\v S_i={1\over 2}\sum_{a}c^\dagger_{i,a,s}\vec{\s}_{ss'}c_{i,a,s'}$, where $a=d_{z^2},d_{xz},d_{yz},d_{xy},d_{x^2-y^2}$ is the orbital index. The last term of \Eq{eff}, with $V>0$,  describes the tendency toward the $d_{xz}/d_{yz}$ orbital ordering\cite{yi,ku,Phillips}. 
 We wrote the last two  terms of \Eq{eff} in real space, but they should be understood as been projected to the band eigen bases that lie in a thin shell around the fermi surface.

\Eq{eff} describes several competing (or ``intertwined'') instabilities, a hallmark of strongly correlated systems. 
Of course solving the effective Hamiltonian is a difficult problem. In the following we shall assume Cooper pairing is the winning instability, and our goal is simply to determine which pairing symmetry is favored the most.
Some technical points: (1) In \Eq{eff} we allowed  the magnetic interaction to extend over arbitrary neighbors. However in Ref.\cite{zhai} it has been shown if one retains only the first ($J_1$) and second neighbor ($J_2$) interaction, the effective Hamiltonian already qualitatively captures the numerical functional renormalization group results for the $\v k$ dependence of the AFM and SC order parameters. Since the purpose of this paper is to elucidate qualitative physics rather than providing quantitative  predictions, we shall truncate $J_{ij}$ to only $J_1$ and $J_2$\cite{hu}.   (2) Once the fermi surface is fixed, the second term of \Eq{eff} has no effect on Cooper pairing.
  Hence we shall drop it in the subsequent discussions. 
Finally on the semantics: when the various interaction in \Eq{eff} fail to drive long range order, we call them ``fluctuations''. For example the $J_1,J_2$ terms will be termed ``magnetic fluctuation''.  \\
\begin{figure}
\begin{center}
(a)\includegraphics[scale=.2]{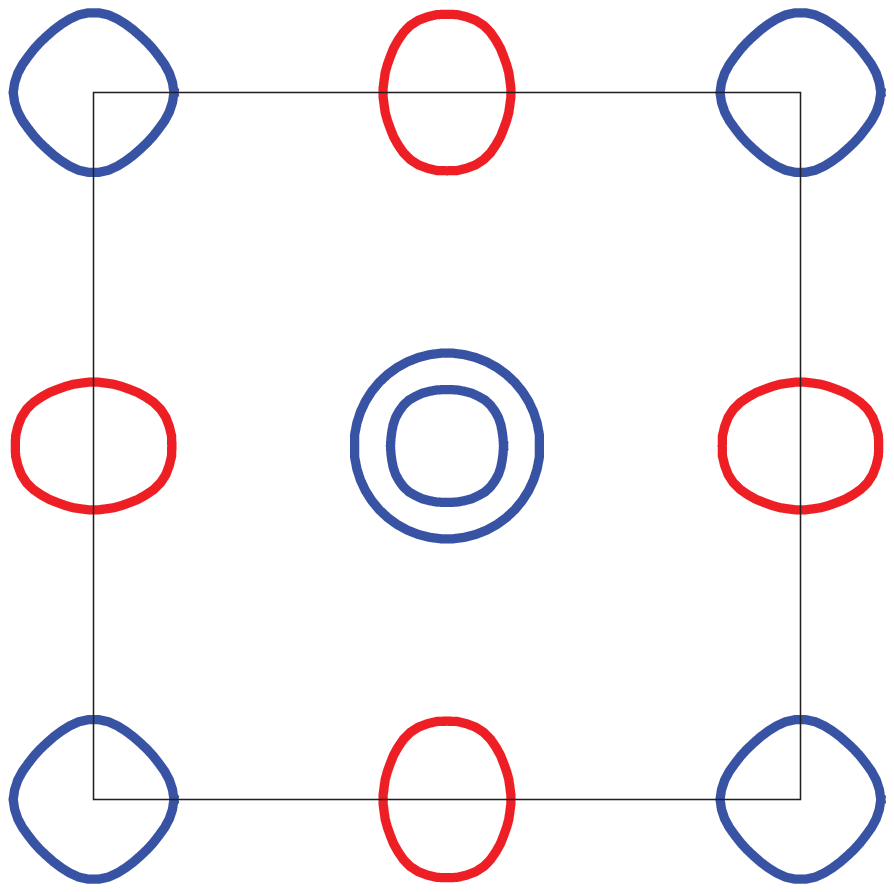}(b)\includegraphics[scale=.2]{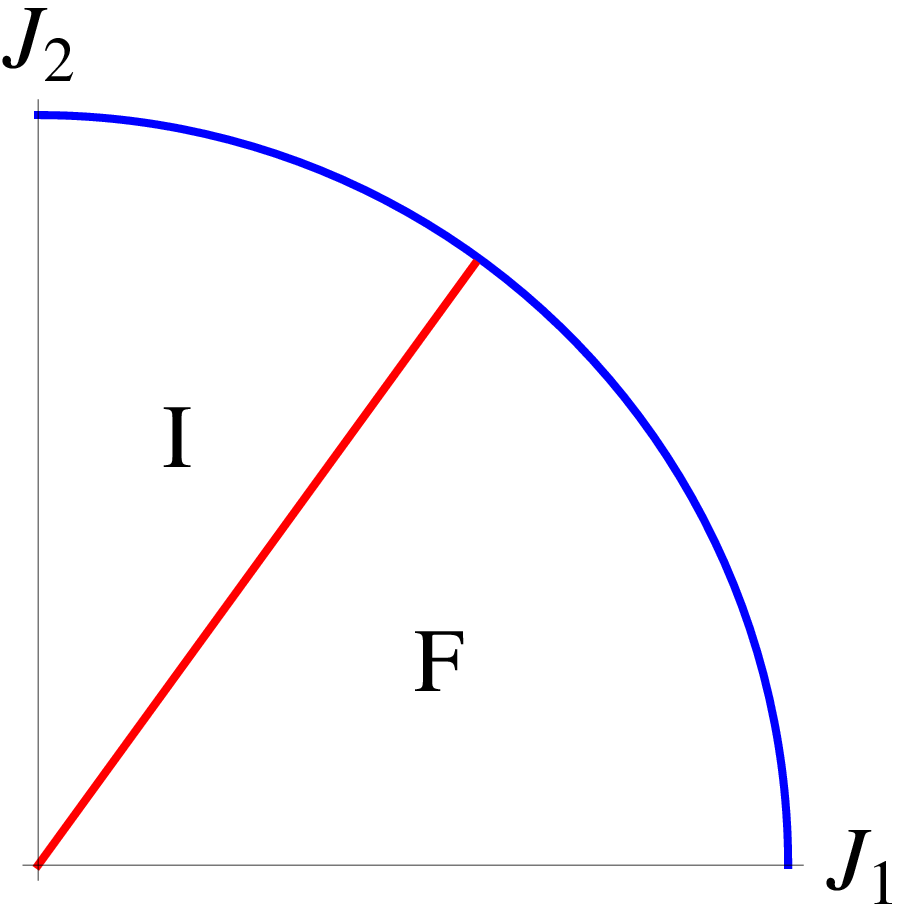}(c)\includegraphics[scale=.5]{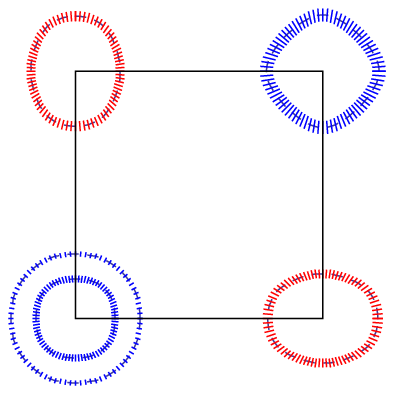}
\caption{(a) A caricature of the fermi surface of Fe-pnictides. Here blue and red mark the hole and electron pockets, respectively. (b) The two regimes of pairing. The value of the critical angle (marked by the red line) is approximately $ 0.3\pi$. In the region marked ``F'' pairing is frustrated. (c) The typical gap function in region ``I'' of panel (b). There is hatch size is proportional to the magnitude and the color indicate the sign (blue: plus and red: minus). Only the first quadrant of the unfolded Brillouin zone is shown. }
\label{1}
\end{center}
\end{figure}

{\bf Results for Fe-pnictide:} A plot of the typical fermi surface for the Fe-pnictide systems is shown in \Fig{1}(a), where both electron pocket (marked red) and hole-pocket (marked blue) are present. First we study the effect of magnetic interaction on pairing. We control the ratio between the AF $J_1$ and $J_2$  by introducing an angle $\theta$
\be
J_1=J\cos\theta,~~J_2=J\sin\theta,~~0\le\theta\le\pi/2.\ee We then project the effective Hamiltonian onto the singlet pairing channel
to construct the following pairing ``matrix''
\be M(\alpha,\phi;\beta,\phi')=U(\alpha,\phi;\beta,\phi'){k_{F\beta}(\phi')\over v_{F\beta,r}(\phi')}.
\label{pairm}
\ee
Here $U(\alpha,\phi;\beta,\phi')$ is the effective singlet pairing interaction on the fermi surface ($\alpha$ labels fermi pockets and $\phi$ is the
angle around them), $k_{F\beta}$ and $v_{F\beta,r}$ are the magnitudes of the fermi wavevector and the projection of the fermi velocity along
the radial direction, respectively.
The leading gap function is the eigenfunction of $M(\alpha,\phi;\beta,\phi')$ with the most negative eigenvalue.  In case  $ M(\alpha,\phi;\beta,\phi')$ has degenerate eigenvalues non-quadratic interaction terms between different superconducting order parameters are necessary to determine the pairing symmetry\cite{WuCJ-tbreaking,Tesanovic-tbreaking}. In this paper we shall just focus on the quadratic instability. Depending on the value of $\theta$, there exists two pairing regimes marked by ``I'' and ``F'' in \Fig{1}(b). The representative gap functions in  region I is shown in \Fig{1}(c), it is s$\pm$.
\begin{figure}
(a)\includegraphics[scale=0.8]{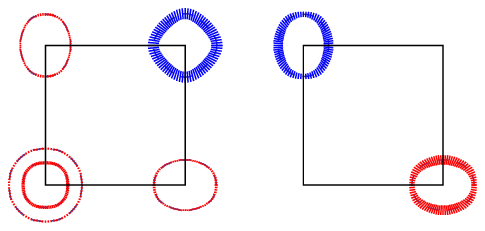}(b)\includegraphics[scale=0.8]{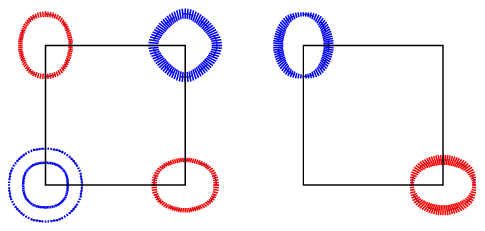}
\caption{The gap function in region ``F'' of \Fig{1}(b). In the presence of the hole pockets (left panels in (a) and (b)) the gap function has s-wave symmetry and the sign of the gap function on the hole pockets changes as $\theta$ increases from 0 (Fig.2(a)) to 0.3 $\pi$ (Fig.2(b)). After removing the hole pockets (right panels in (a) and (b)) the symmetry becomes d-wave.  Only the first quadrant of the unfolded Brillouin zone is shown.}
\label{2}
\end{figure}
Cooper pairing in this region is {\it non-frustrated} in the sense that if we remove the hole pockets and recalculate the gap function on the electron pockets, the same sign structure is obtained.

In contrast, in region F pairing is frustrated. In \Fig{2}(a,b) we show the gap function with and without the hole pockets in this region. Generally speaking the pairing symmetry is s-wave with hole pockets and d-wave without.
\begin{figure}
\begin{center}
(a)\includegraphics[scale=.25]{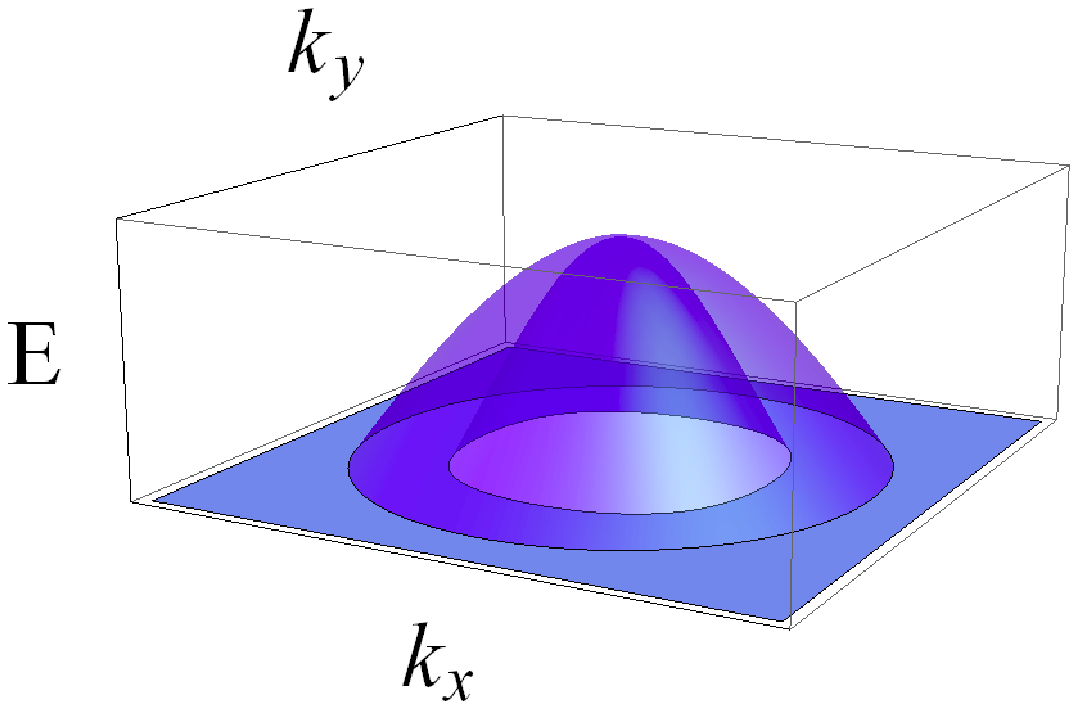}(b)\includegraphics[scale=.25]{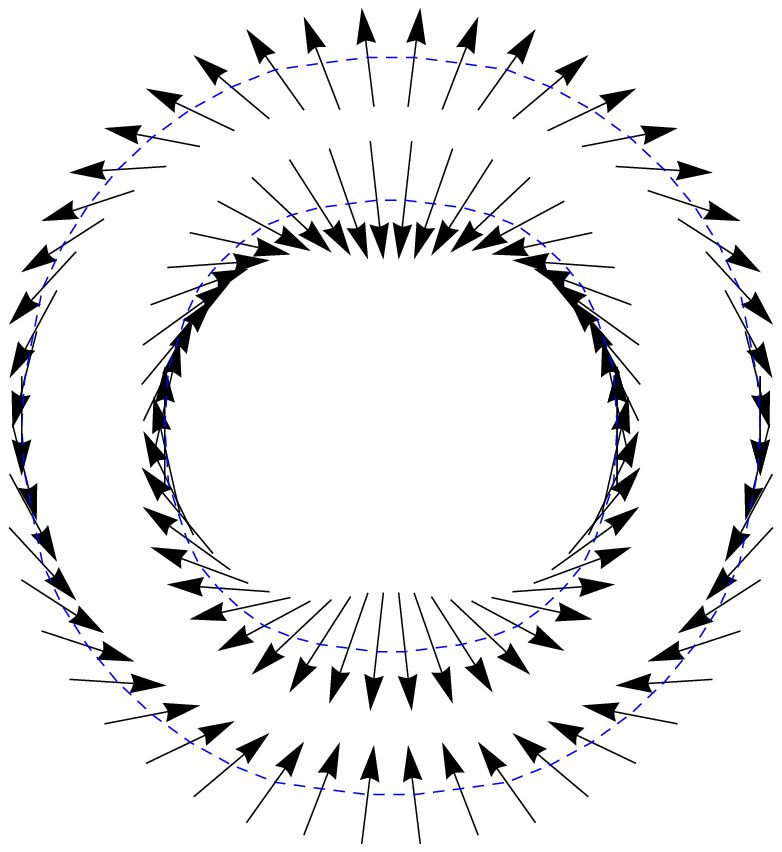}(c)\includegraphics[scale=.25]{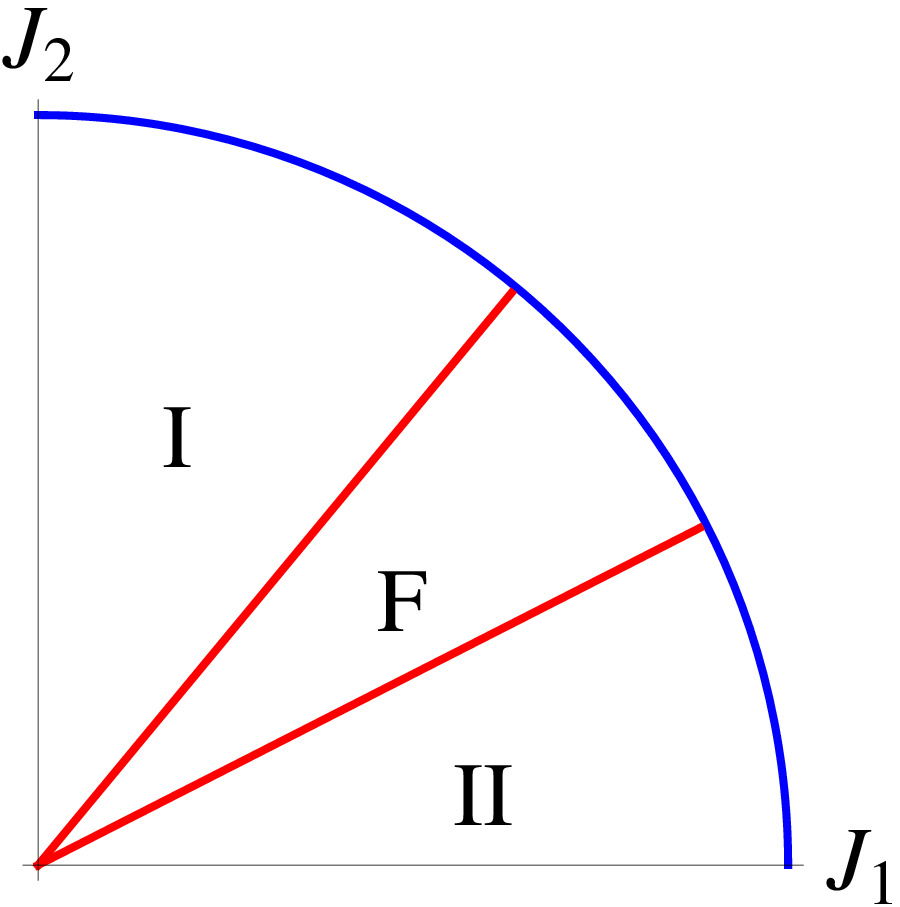}
\caption{The band vorticity associated with the hole pockets. (a) The band structure near $\Gamma$. (b) The pseudo spin winding  (see text).(c) The region with pairing frustration substantially shrinks after the removal of the band vorticity of the hole Fermis surfaces. The values of the two critical angles (marked by the red lines) are approximately $0.15\pi$ and $0.28\pi$. Only the first quadrant of the unfolded Brillouin zone is shown.}
\label{3}
\end{center}
\end{figure}
The fact that after removing the hole pockets the symmetry of the gap function changes is reminiscent to the change of spin alignment from coplanar to collinear after removing a spin to unfrustrate the AF Heisenberg interaction on a triangle. In the literature\cite{scalapino,SiQM} it is noted that there is a near degeneracy between the $s_{\pm}$ and $d_{x^2-y^2}$ pairing symmetry for values of the interaction parameters
yielding significant pairing strength. We claim this degeneracy is due to the pairing frustration discussed above.

It turns out the pairing frustration is, to a large extent, caused by the band topology of the hole pockets. In \Fig{3}(a) we show the dispersion of the hole bands near the center of the Brillouin zone. The double degeneracy at $\Gamma$ is protected by the point group symmetry. A consequence of such degeneracy is the existence of ``band vorticity'' at $\Gamma$ \cite{QiXL-vorticity,RanY-vorticity}. To see that we first note the orbital content of the hole band eigenfunctions are predominantly $d_{xz}$ and $d_{yz}$. Let's use a pseudo spin 1/2 to represent these two orbital states, e.g., $d_{xz}\ra \tau_z=1/2$ and $d_{yz}\ra \tau_z=-1/2$. Using the band eigenfunctions we compute the expectation value of the pseudospin. The result lies in the x-z plane and the direction winds around the fermi surface (with vorticity 2) as shown in \Fig{3}(b).

\begin{figure}
\begin{center}
(a) \includegraphics[scale=.8]{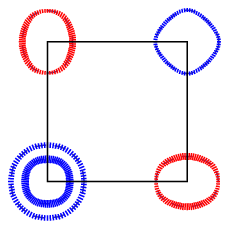} (b) \includegraphics[scale=.8]{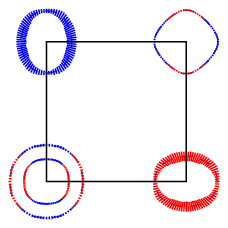} (c)\includegraphics[scale=.8]{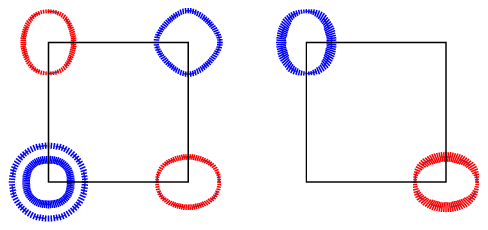}
\caption{(a) Typical gap function in region ``I'' of \Fig{3}(c), there is no pairing frustration. (b) Typical gap function in region ``II'' of \Fig{3}(c), there is no pairing frustration. (c) Typical gap function in region ``F'' of \Fig{3}(c). There is pairing frustration. The left and right panels are the gap functions with and without the hole pockets. Only the first quadrant of the unfolded Brillouin zone is shown.}
\label{4}
\end{center}
\end{figure}

After we remove the hole band vorticity (by modifying the band wavefunction while maintaining the band dispersion) the frustrated pairing region substantially  shrinks, as shown in \Fig{3}(c). The typical gap function in region ``I'',``II'' and ``F'' are shown in \Fig{4}(a,b,c). From this we conclude, the band vorticity associated with the hole pockets is an important cause of the pairing frustration. It is important to note, however, removing the band vorticity does not completely eliminate pairing frustration. What's needed is the removal of the hole pockets entirely!

This motivates us to make the following conjecture: one reason that the absence of hole pockets raises $T_c$ in A$_x$Fe$_{2-y}$Se$_2$ and FeSe/STO is because the pairing frustration has been removed. However high $T_c$ also requires strong AFM fluctuation (at least in the magnetic pairing scenario).  If the AFM correlation is due to the nesting between the electron and hole fermi surfaces\cite{mazin}, one would predict the weakening of AFM fluctuation (hence the weakening of pairing). However the recent neutron scattering\cite{jun} experiment performed on ``semiconducting'' K$_x$Fe$_{2-y}$Se$_2$ compounds neighboring the superconducting ones has revealed strong stripy AFM long range order peaking at the same wave vector as that observed in the parent pnictide compounds. This result suggests that it is better to view the AFM correlation in FeSC as due to local correlation physics, similar to the cuprates, instead of as coming from fermi surface nesting. More importantly it tells us the AFM fluctuation, needed for Cooper pairing, is strong in K$_x$Fe$_{2-y}$Se$_2$ despite the absence of the hole pockets.

\begin{figure}
\begin{center}
(a)\includegraphics[scale=0.25]{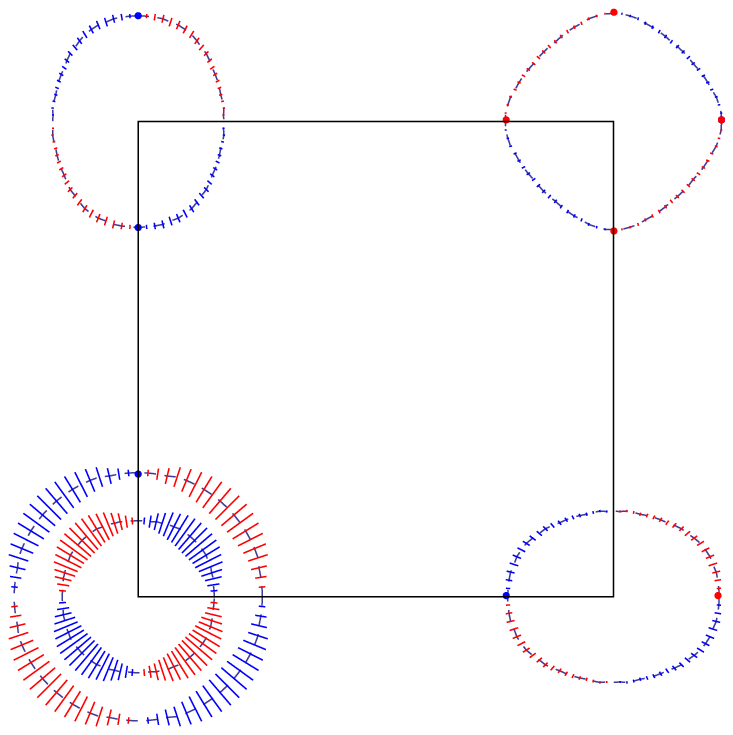}(b)\includegraphics[scale=0.9]{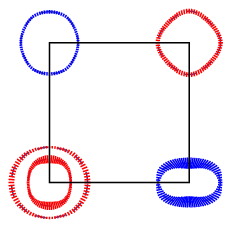}(c)\includegraphics[scale=0.9]{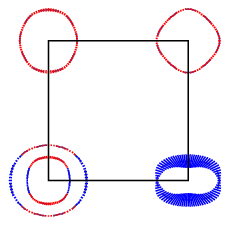}
\caption{(a) The gap function that is suppressed by the orbital fluctuation. (b,c) In the presence of static orbital ordering the frustration region also shrinks compared to \Fig{1}(b). The phase diagram is qualitatively similar to \Fig{3}(c) except the ``F'' region is slightly wider. The gap function in region ``I'' and ``II'' are shown in (b) and (c).Only the first quadrant of the unfolded Brillouin zone is shown. }
\label{5}
\end{center}
\end{figure}

Motivated by Ref.\cite{chu2} we next ask what is the effect of orbital fluctuation  on Cooper pairing.
 To mimic orbital fluctuation we turn on the $V$ term in \Eq{eff}. To our surprise, for all values of $V$ this has no effect on pairing. 
To understand this result let us switch off $J_1$ and $J_2$ and look at the eigenfunctions of \Eq{pairm} when there is only the orbital interaction.
It turns out in this case except one positive eigenvalue all other eigenvalues of $M$ are zero. In \Fig{5}(a) we illustrate the eigenfunction associated with the
positive eigenvalue.  Thus the orbital interaction lifts up the pairing eigenvalue associated the $d_{xy}$ symmetry but leaves all other symmetry channels unaffected. In particular, this lifting does not affect the leading pairing symmetries, namely $s$ and $d_{x^2-y^2}$ in \Fig{1} and \Fig{2}. The fact that $V$ introduces  only a single positive pairing eigenvalue is due to the on-site nature of the orbital interaction. Had we allowed for further neighbor orbital interaction, e.g.,
$V_0\sum_i n_{i,xz}n_{iyz}+V_1\sum_{\langle ij\rangle} n_{i,xz}n_{j,yz}$ there will be more than one positive eigenvalues. For example if we choose  $V_1/V_0=0.5$ there are five positive eigenvalues (all other eigenvalues are still zero). Again, none of these five eigenfunctions has the symmetry of the leading pairing channel in \Fig{1} and \Fig{2}. For all the cases we studied this is always true. Thus we conclude orbital fluctuation has no effect on pairing.

Because Ref.\cite{matzuda} claimed the superconducting dome in BaFe$_2$(As$_{1-x}$P$_x$)$_2$ is embedded in a non-magnetic nematic phase, we next study the effect of static orbital order on pairing. In \Fig{5}(b,c) we show the anisotropic gap function in the presence of a $\delta E=40$ meV energy splitting between the xz and yz orbitals, namely, $\delta E\sum_i (n_{i,xz}-n_{i,yz})$.
If this anisotropy is observed by ARPES it will constitute a strong evidence for coexistence of superconductivity and nematic order.\\

\begin{figure}
\begin{center}
(a)\includegraphics[scale=.25]{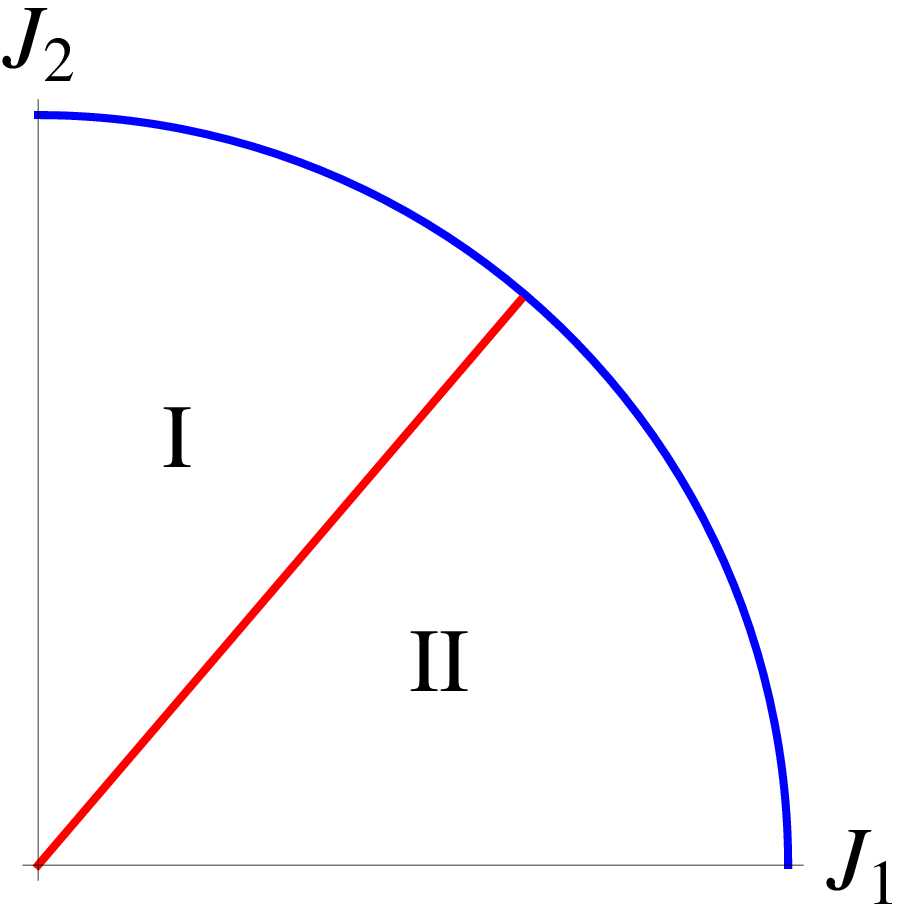}(b)\includegraphics[scale=.3]{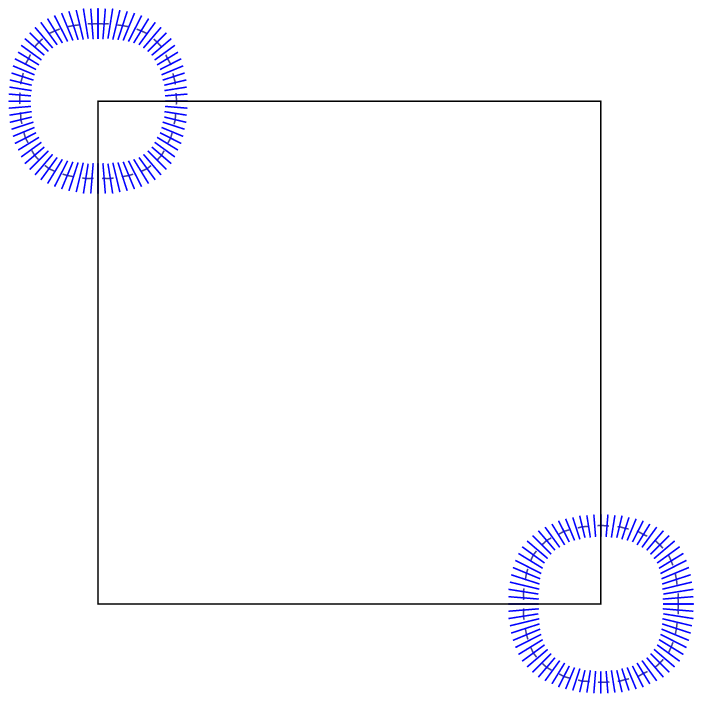}(c)\includegraphics[scale=.3]{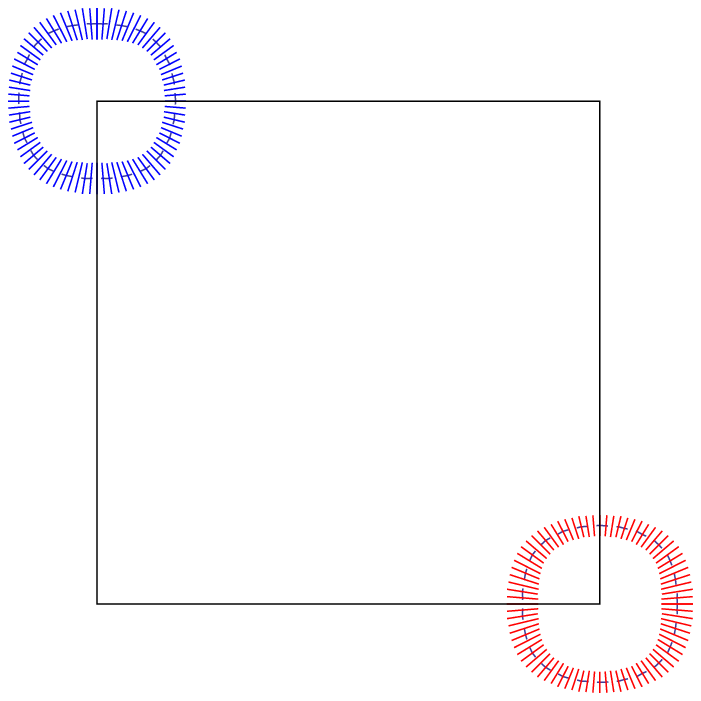}
\caption{Magnetic fluctuation induced pairing symmetry of systems with only electron Fermi surfaces. (a) Two different pairing regimes. The value of the critical angle (marked by the red line) is approximately $0.27\pi$. The gap function in region I (b) and II (c). Only the first quadrant of the unfolded Brillouin zone is shown.}
\label{7}
\end{center}
\end{figure}

{\bf Results for the Fe-chalcognide:} The fermiology of superconducting A$_x$Fe$_{2-y}$Se$_2$ and FeSe/STO are rather similar - there are only electron pockets.
As mentioned earlier due to the neutron scattering result of Ref.\cite{jun} there is every reason to believe the same magnetic interaction exists in the effective Hamiltonian (at least for K$_x$Fe$_{2-y}$Se$_2$ ). Using bandstructures with only electron pockets\cite{faepl,fese} our result for the gap function under different $J_1$ and $J_2$ ratio is shown in \Fig{7}. Depending on whether $J_1$ or $J_2$ dominates, the pairing symmetry is either $s$ wave or fully gapped $d$-wave, consistent with the findings in Ref.\cite{Thomale}. Interestingly despite the fact that AFM effective interaction is characteristic of a repulsive system, the in-phase s-wave pairing is found for a large region of parameter space.

Because nodeless $d$-wave occupies a significant region of parameter space in \Fig{7}(a)),  it is important to investigate the effect of the electron pocket hybridization on the gap function. This hybridization exists (even for $k_z=0$) when the the $z\ra -z$ glide plane symmetry is broken. For A$_x$Fe$_{2-y}$Se$_2$
this occurs near the sample surface (which is what ARPES probes). For FeSe/STO there is no glide plane symmetry at all.  It was first pointed out in Ref.\cite{mazin2} that when such hybridization is sufficiently strong it can modify the nodeless d-wave pairing state  into nodal $d$-wave pairing.

\begin{figure}
\begin{center}
(a)\includegraphics[scale=.3]{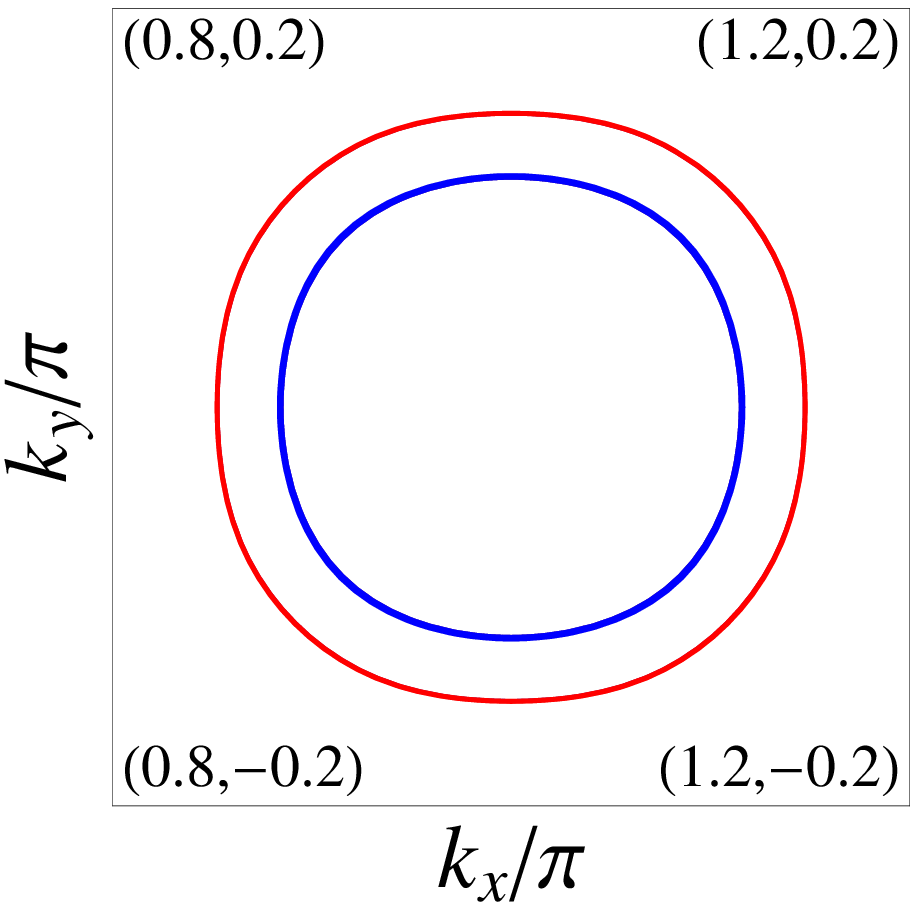}~~~(b)\includegraphics[scale=.35]{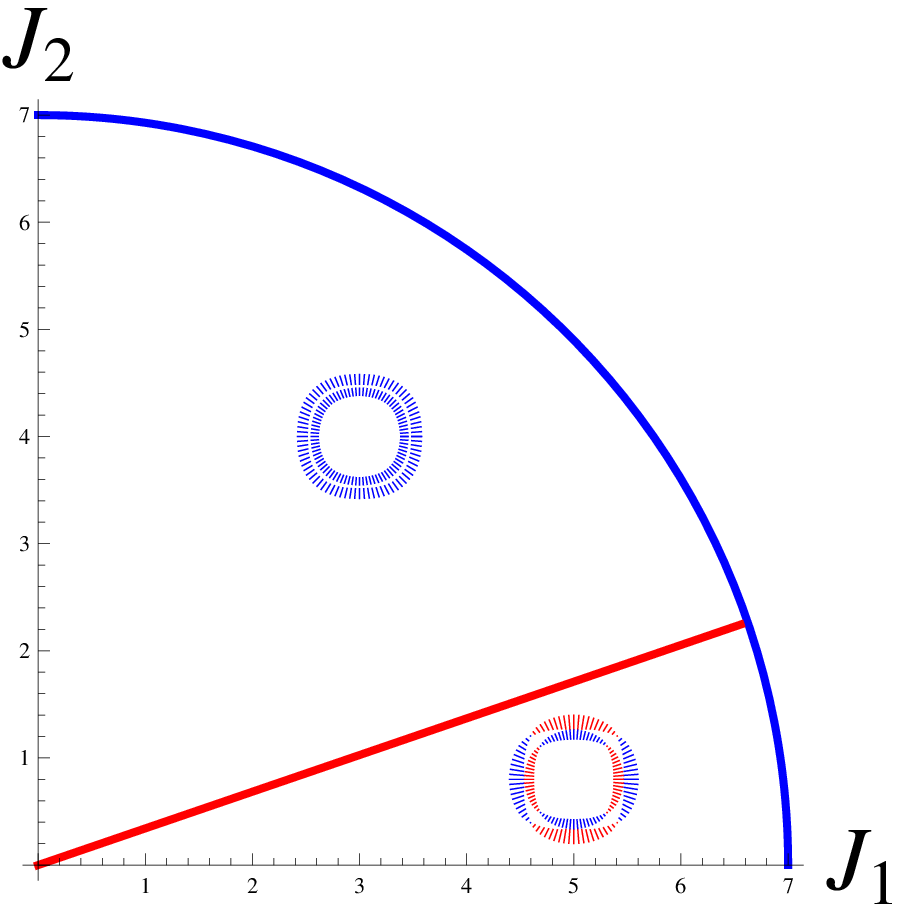}
\caption{(a) The hybridization-split electron pockets. (b) The gap function in the presence of hybridization. In panel (b) the splitting between the two electron fermi surfaces has been enhanced to increase clarity. The value of the critical angle (marked by the red line) is approximately $0.105\pi$.}
\label{8}
\end{center}
\end{figure}

In \Fig{8}(a) we choose the hybridization matrix elements so that electron pockets is only slightly split and remain more or less circular. This choice is motivated by the fact that ARPES found rounded electron pocket without discernable splitting for both  A$_x$Fe$_{2-y}$Se$_2$\cite{donglai} and FeSe/STO\cite{xj} .

In the presence of the electron pocket splitting, the phase diagram and gap function are shown in \Fig{8}(b). The hybridization has clearly increased the range of stability for the $s$-wave pairing. Indeed there is a region of the phase diagram where the nodeless d-wave pairing has turned into a nodal one. However, the dominant pairing symmetry in \Fig{8}(b) is the in phase $s$-wave rather than the out-of-phase $s$-wave proposed in Ref.\cite{mazin2,chubukov}. This result is consistent with the recent ARPES result which rules out d-wave pairing based on the mapping of the gap function on the electron pockets center around $\v k=(0,0,\pi)$\cite{donglai2}.

Finally we believe the substrate, namely SrTiO$_3$, plays an important role in raising the $T_c$ of FeSe/STO and further stabilize s-wave pairing.  This is through the screening of the long range part of the Coulomb interaction\cite{fese}. As far as orbital fluctuation and orbital ordering are concerned, the results are very similar to those for systems with both electron and hole pockets.
\\

{\bf Conclusion:} We have
used an effective Hamiltonian to address two important modern issues raised by the high T$_c$ A$_x$Fe$_{2-y}$Se$_2$ and FeSe/STO superconductors.
This effective Hamiltonian is constructed under the guidance of (1) FRG calculation\cite{zhai}, (2) VMC calculation\cite{fan} and (3) experiments\cite{afm1,afm2,afm3,hosono,zhao,davis,yi,chu1,chu2,matzuda}.  The issues are (1) the effect of fermiology on high temperature superconductivity and (2) the effects of orbital ordering and orbital fluctuations on Cooper pairing. Our conclusions are summarized in the third paragraph of the introduction.\\

{\bf Acknowledgments:} We Acknowledge Jun Zhao, Bob Birgeneau, Alessadra Lanzara and Qiang-hua Wang for helpful discussions. DHL acknowledges the support by the DOE grant
number DE-AC02-05CH11231. FY is supported by the NSFC under the grant number 11274041, and the NCET program under the grant number NCET-12-0038.

\end{document}